\documentclass[pra,aps,showpacs,floatfix,twocolumn]{revtex4-1}
\usepackage{graphicx}
\newcommand{\be}{\begin{eqnarray}}
\newcommand{\ee}{\end{eqnarray}}
\def\beq{\begin{equation}}
\def\eeq{\end{equation}}

\begin{document}
\title{Stability of the superfluid state in a disordered 1D ultracold fermionic gas}
\author{Masaki Tezuka}
\altaffiliation
[Present address: ]
{Department of Physics, Kyoto University, Kitashirakawa, Sakyo-ku, Kyoto 606-8502, Japan}
\affiliation{Department of Physics, University of Tokyo, Hongo, Bunkyo-ku, Tokyo 113-0033, Japan}
\email{tezuka@scphys.kyoto-u.ac.jp}
\author{Antonio M. Garc\'{\i}a-Garc\'{\i}a}
\affiliation{CFIF, Instituto Superior T\'ecnico, 
Universidade T\'ecnica de Lisboa, Av. Rovisco Pais, 1049-001 Lisboa, Portugal}
\begin{abstract}
We study a 1D Fermi gas with attractive short range-interactions in a disordered potential by the density matrix renormalization group (DMRG) technique. Our results can be tested experimentally by using cold atom techniques. We identify a region of parameters for which disorder enhances the superfluid state. As disorder is further increased, global superfluidity eventually breaks down. However this transition seems to occur before the transition to the insulator state takes place. This suggests the existence of an intermediate metallic `pseudogap' phase characterized by strong pairing but no quasi long-range-order. 
 \end{abstract}
\maketitle
\section{Introduction}
It is now possible to realize experimentally disorder and interactions with unprecedented precision
by using cold atom techniques \cite{aspect,fallani}. This is an ideal setting to test theoretical predictions on novel phases of quantum matter and quantum phase transitions \cite{condis}.     
Motivated by these possibilities we study a disordered 1D Fermi gas with short-range attractive interactions by the DMRG technique. The effect of disorder is mimicked by a quasiperiodic (multichromatic) potential. Both the potential \cite{fallani} and the interaction can be implemented experimentally.
Our main results can be summarized as follows:
a) attractive interactions enhance localization effects. The critical disorder at which the metal-insulator transition occurs decreases as the interaction becomes stronger; b) in contrast to higher dimensions, fluctuations in the metallic phase, but close to the insulator transition, break down quasi long-range order. The resulting anomalous metallic region has 'pseudo-gap' features; c) in the superfluid phase, and for moderate interactions, disorder enhances quasi long-range order. 
\\
We start with a brief overview of previous research on this problem.
In the non-interacting limit the nature of the eigenstates of a 1D tight-binding model, with hopping $t\equiv 1$, in the quasiperiodic potential \cite{fallani}
\be
\label{qp}
V(n) = \lambda \cos(2\pi \omega n +\theta)
\ee
with $\omega$ irrational and $\theta \in [0,2\pi)$ depends on the value of the disorder strength $\lambda > 0$.
All the eigenstates are exponentially localized \cite{jito,komo} for $\lambda > 2$ with a localization length $\propto 1/|\lambda -2|$.
For $\lambda < 2$ the quantum dynamics is similar to that of a free particle in a periodic potential. For $\lambda = 2$ the system undergoes a metal-insulator transition \cite{komo}. We note that the potential is strongly correlated $\langle V(n)V(0) \rangle \propto \cos(2\pi \omega n)$ \cite{fish1}. In 1D, a non-decaying $\langle V(n)V(0) \rangle$  is a necessary condition \cite{kotani} for the existence of a band of metallic states.   
In the limit  $\lambda \to 0$ an exact solution for a continuous 1D model with short range attractive interactions -- the Gaudin - Yang model \cite{gaudin} --
is available \cite{fuchs,gaudin,lieb}. An exact solution is also known for the discrete version of this model, the 1D Hubbard model \cite{hubbard,korepin}.
For $|U| \ll 1$ pairing is BCS-like. For $|U|\to \infty$ the system behaves as a hard-core Bose gas \cite{tonk}. \\
It was found in \cite{shulz} that the addition of a weak Gaussian disorder induces a metal-insulator transition for sufficiently strong interactions. The effect of a quasiperiodic potential has also been addressed in the literature \cite{hiro, schreiber,chavez,giamarchi}. 
The numerical results of \cite{hiro} indicate that the critical disorder at which the metal-insulator transition occurs depends on the strength of the interaction. By contrast the DMRG analysis of \cite{schreiber} concluded that, for spinless fermions, the critical disorder is the same as in the non-interacting case. In \cite{shiro}, also employing a DMRG technique, 
it was found that the presence of a weak disordered potential enhances superfluidity. 
 
Bosonization techniques combined with a renormalization group analysis were employed in \cite{giamarchi} to investigate the effect of interactions in another 1D quasiperiodic system, the Fibonacci chain \cite{koh1}. 
The perturbative treatment of \cite{giamarchi} showed 
that the critical disorder depends on both the strength of the interactions and the position of the Fermi level. We note that, as in Eq.(\ref{qp}), correlations of the potential studied in \cite{giamarchi} are very strong $\lim_{n \to \infty} \langle V(n)V(0) \rangle \neq 0$. However different reasons prevent from a direct comparison between these models: a) in \cite{giamarchi} only spinless fermions are considered, b) for $\lambda \ll 1$, the limit in which the formalism of \cite{giamarchi} is applicable, our system is in the metallic region with properties almost identical as those of a periodic potential, c) for no interactions the spectrum of the Fibonacci chain is singular continuous for all $\lambda$. This leads to eigenstates that are power-law localized and quantum superdiffusion \cite{koh1}. Such features are only found in Eq.(\ref{qp}) for $\lambda = 2$. For $\lambda < 2$ the spectrum of Eq.(\ref{qp}) is absolutely continuous as in a perfect metal. 
 
For results on the dynamics of a Bose gas in a quasiperiodic potential we refer to \cite{coldbos1D}.
Mean field approaches in 1D are problematic since fluctuations, specially in the presence of a disordered potential, are not negligible. We thus anticipate qualitative differences with respect to the 2D and 3D cases where, for disorder weak enough, quasi long-range order persists \cite{gosh} even in the insulator region provided that the localization length is larger than the coherence length. Finally we mention that the effect of disorder in
Fermi gases of higher dimensions has been investigated in \cite{orso} using mean field techniques and neglecting Anderson localization effects \cite{anderson}. For numerical studies on the attractive Hubbard model in a disordered potential we refer to \cite{scal}.
\section{The model}
We study the discrete $L$-site Hubbard model,
\begin{eqnarray}
\hat H &=& -\sum_{i=1,\sigma}^{L-1}(\hat c_{i-1,\sigma}^\dag
\hat c_{i,\sigma}+\mathrm{h.c.})
+ U \sum_{i=0}^{L-1} \hat n_{i,\uparrow}\hat n_{i,\downarrow}\nonumber\\
&+& \sum_{i=0}^{L-1} V(i) \hat n_{i},
\label{eqn:Hubbard}
\end{eqnarray}
where 
$\hat c_{i,\sigma}$ annihilates an atom at site $i$ in spin state $\sigma(=\uparrow, \downarrow)$,
$\hat n_{i,\sigma}\equiv \hat c_{i,\sigma}^\dag \hat c_{i,\sigma}$,
$\hat n_i\equiv \hat n_{i,\uparrow}+\hat n_{i,\downarrow}$,
 $V(i)$ is given by Eq.(\ref{qp}) with
$\omega\equiv F_{n-1}/F_n$ the ratio of two consecutive Fibonacci numbers, $L=F_n+1$ and $\theta=0$ so that $V(0)=V(L-1)=\lambda$.
We note that we have set the hopping integral $t \equiv 1$.
\begin{figure}[h]
\includegraphics[width=0.82\columnwidth,clip,angle=0]{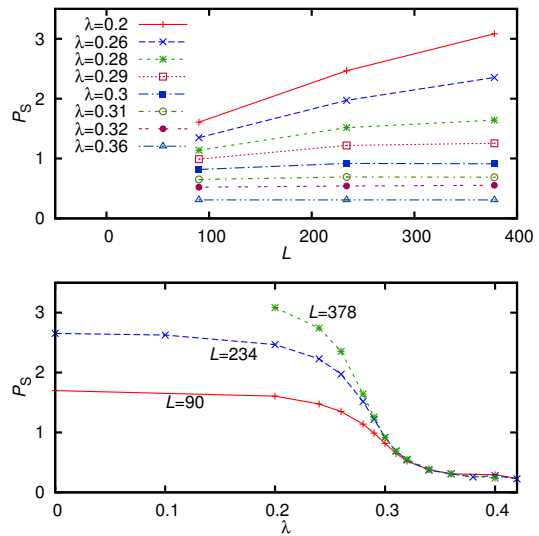}
\caption{(Color online) Upper: Pairing structure factor $P_{\mathrm{s}}$ Eq.(\ref{ps}), as a function of the system size $L$ for different $\lambda$'s and $U = -6$. Superfluidity, characterized by an increasing $P_s(L)$, is observed up to $\lambda_c \approx 0.29$. Lower: $P_\mathrm{s}$ as a function of disorder also for $U = -6$ and different sizes. A $P_{\mathrm{s}}$ almost independent of $L$ is a signature of broken quasi long-range order. See text for more details.}
\label{stiffa}
\end{figure}

\begin{figure}
\includegraphics[width=0.82\columnwidth,clip,angle=0]{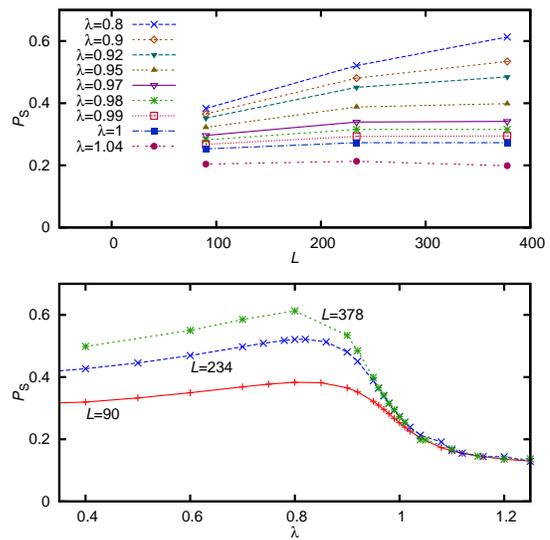}
\caption{(Color online) Same as Fig.\ref{stiffa} but for $U = -1$. Upper: $P_\mathrm{s}$ only increases with $L$ for $\lambda \lesssim 0.95$. Global superfluidity is thus broken at $\lambda_{\mathrm{c}} \approx 0.95$. Lower: $P_{\mathrm{s}}$ is an increasing function of $\lambda$ until $\lambda \approx 0.8$. Therefore the quasiperiodic potential enhances superfluidity for moderate disorder.}
\label{stiffb}
\end{figure}
\begin{figure}[pbt]
\includegraphics[width=0.82\columnwidth,clip,angle=0]{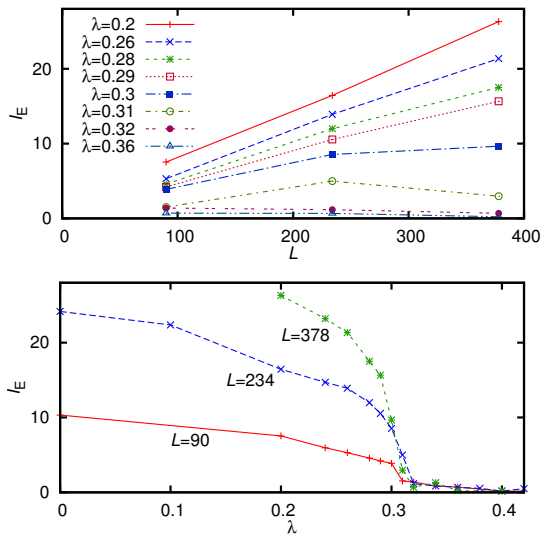}
\caption{(Color online) Upper: $I_E$, Eq.(\ref{ipr}) as a function of $L$, for different $\lambda$'s and $U = -6$. A metal-insulator transition
is observed at $\lambda^\mathrm{ins}_{\mathrm{c}} \approx 0.31$. However for $U = 0$, $\lambda^\mathrm{ins}_{\mathrm{c}} = 2$. Therefore attractive interactions enhance localization. Lower: $I_E$ as a function of $\lambda$. An increase of $I_E$ with the system size is a signature of a metal.}
\label{locali}
\end{figure}
\begin{figure}[pbt]
\includegraphics[width=0.82\columnwidth,clip,angle=0]{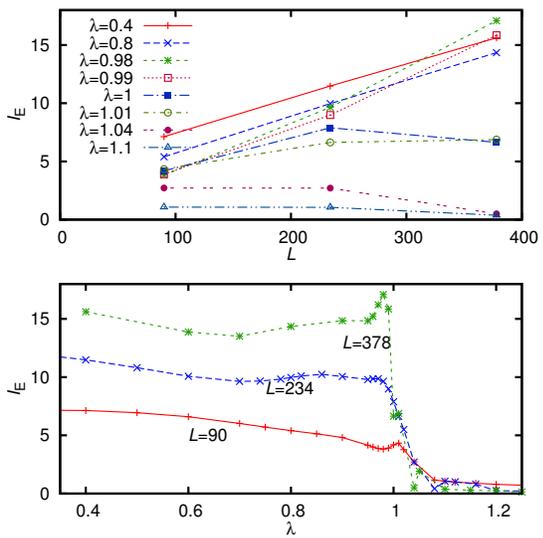}
\caption{(Color online) Same as Fig.\ref{locali} but for $U = -1$. Upper: The metallic state is characterized by a $I_E$ that increases with $L$. The insulator transition occurs at $\lambda^\mathrm{ins}_{\mathrm{c}} \approx 1.0$. In contrast to the $U = -6$ case, it is observed a further increase of $I_E$ very close to the transition $\lambda \approx 0.99$. This is a consequence of the enhanced eigenfunction correlations in this region \cite{mirlin,condis}. For $U = -6$ the coherence length is much smaller and consequently eigenfunctions correlations are suppressed. Lower: Also, in contrast with the $U = -6$ case, the metallic state is also enhanced for intermediate disorder $\lambda$'s below the transition. This is again a quantum coherence effect caused by the bands structure of the quasiperiodic potential (see text).}
\label{locali1}
\end{figure}
\begin{figure}
\begin{center}
\includegraphics[width=0.82\columnwidth,clip,angle=0]{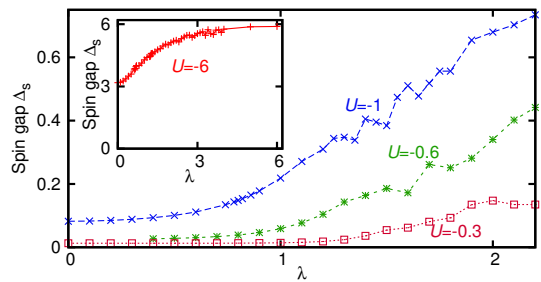}
\end{center}
\caption{(Color online) Spin gap $\Delta_{\mathrm{S}}$, Eq.(\ref{deltas}) as a function of $\lambda$ for different $U$'s and fixed $L$. For small $\lambda$ the gap is an increasing function of disorder as a consequence of the band structure of the quasiperiodic potential. Close to the insulator transition $\lambda \lesssim \lambda_c$ there is an additional gap enhancement caused by eigenfunction correlations \cite{condis}.}
\label{spingap}
\end{figure}
The behavior of Hamiltonian Eq.(\ref{eqn:Hubbard}) in certain limits is already known:
a) for $|U| \gg 1$ the system maps onto a weakly interacting bosonic gas with
a kinetic term which is $1/|U|$ smaller than in the original fermionic model. Therefore the critical disorder at which the transition to localization occurs is $\lambda_ c \approx 2/|U|$ \cite{shepe}; b) the coherence length for weak disorder ($\lambda \ll \lambda_{\mathrm{c}}$) is $\xi_\mathrm{co} \propto 1/U^2$ for $|U| \gg 1$ and $\xi_\mathrm{co} \propto e^{1/|U|}$ for $|U| \ll 1$; 
c) for $U \lesssim 1$ not very large, the spin gap (see Eq.(\ref{deltas}))  $\Delta_\mathrm{S} \propto 1/\xi_\mathrm{loc}$, \cite{gosh} with $\xi_\mathrm{loc}$ the localization length.\\
The above information is enough to put forward a tentative description of the system phase diagram (in the $U < 0$, $\lambda$ plane):
a) for fixed $|U| \gg 1$ and $|U| \ll 1$, the loss of quasi long-range order and
the transition to the insulator phase will occur at similar $\lambda$'s: $\lambda_c \approx 2/|U|$ and $\lambda_c \approx 2$ respectively;
b) for intermediate $U$ it might be possible that the two transitions take place at slightly different $\lambda$'s as the
breaking of superfluidity might be induced by phase and amplitude fluctuations in the metallic region. \\
In order to test the validity of these qualitative arguments we study the Hamiltonian Eq.(\ref{eqn:Hubbard}) with the potential Eq.(\ref{qp}) by the DMRG technique.
The filling factor $\nu$ is kept constant $\nu =
N / L =  1/9$ for $(N, L)=(10,90),(26,234),(42,378)$ -- quantitatively our results might depend on $\nu \ll 1$  
\cite{giamarchi} --.
We obtain the ground state for $N$ spin-up and $N$ spin-down atoms.
Up to $m=400$ basis states for each block are kept in the finite-size system DMRG iterations.
\section{Results}
Our first task is to determine for what range of parameters global superfluidity breaks down.
In weakly disordered BCS superconductors a study of the ground state and the low energy excitations is enough to answer this question as
the vanishing of the gap is equivalent to the breaking of global coherence. In strongly disordered and strongly coupled superconductor the situation is different as the gap might be finite even after fluctuations have destroyed global superfluidity \cite{gosh}. It is thus necessary to compute
observables that directly measure the phase stiffness of the system.
\subsection{Phase rigidity: Pairing structure factor}
A popular choice \cite{scal} is the averaged equal-time pairing
structure factor,
\be
\label{ps}
P_{\mathrm{s}} \equiv \left \langle \sum_r \Gamma(i,r) \right \rangle
\ee
where $\langle \ldots \rangle$ stands for the spatial average on the site index $i$, $\Gamma(i,r) \equiv \langle \hat \Delta(i+r) \hat \Delta^\dagger(i) \rangle$ and
 $\hat \Delta(j) \equiv \hat c_{j\uparrow} \hat c_{j\downarrow}$. Quasi long-range order (there is no true order in 1D) occurs for $\Gamma(r) \sim 1/r^K$ for $r \gg 1$. In the case with no disorder it was demonstrated in \cite{korepin}  
that superconductivity correlations are always leading with respect to other types of quantum order and that $K \leq 1$. The limit 
$K = 1$ is only achieved in the limit $|U| \to \infty$. In the disordered case it is also plausible to expect that $K \leq 1$ but $K = 1$ occurs for a finite $U$ which depends on $\lambda$. Therefore we define quasi global superfluidity by $P_s \propto L^{1-K}$ with $K < 1$.  In Figs. \ref{stiffa} and \ref{stiffb} we observe:\\ a) the critical $\lambda = \lambda_c < 2$ at which global superfluidity breaks down decreases as $|U|$ increases. Therefore a tighter binding is correlated with a greater instability to disorder effects \cite{shepe}.\\ 
b) For not too strong $U$, $P_{\mathrm{s}}$ is an increasing function of $\lambda$ 
up to some $\lambda$ close but smaller than $\lambda_{\mathrm{c}}$.\\
c) For $|U| \gg 1$ this feature is not observed.\\
We believe that b) is a coherent effect related to the peculiar band structure induced by the quasiperiodic potential. This is also consistent with c). As $|U|$ increases the coherence length decreases, the details of the spectral density are smoothed out, and no enhancement of superfluidity is observed. 
\subsection{Localization: Density fluctuations} 
 We now turn to localization properties. More specifically we determine numerically the location of the critical disorder $\lambda^\mathrm{ins}_{\mathrm{c}}$ at which the metal-insulator transition occurs.
Different quantities, such as density fluctuations \cite{shepe} or the conductance \cite{schreiber}, provide a similar estimation of localization effects.
However the numerical value of $\lambda^\mathrm{ins}_{\mathrm{c}}$ might depend weakly on the observable employed \cite{carter}.
We present results for the density fluctuations,
\be
\label{ipr}
I_E \equiv \left(\sum_i \delta n(i,N,N)^2\right)^{-1},
\ee
where $\delta n(i,N,N) \equiv n(i,N+1,N+1) - n(i,N,N)$ is the ground-state atomic density at site $i$
for $N$ spin-up and $N$ spin-down atoms, $E$ stands for the ground state energy in this case.
For $U = 0$, it corresponds with the usual definition of the inverse participation
ratio in non-interacting systems \cite{mirlin}.\\
In the insulator region it is proportional to the localization length $I_E \propto \xi_\mathrm{loc}$.
It decreases slowly as disorder increases until it saturates $I_E \to 1/4$ for $\lambda \to \infty$.
In the metallic region ($\lambda \ll \lambda^\mathrm{ins}_{\mathrm{c}}$), $I_E \propto L$ with only a weak dependence on $\lambda$. Close to the critical region, $ I_E \propto L^\alpha$ with $\alpha < 1$ a constant that depends on the eigenstates multifractal
dimensions \cite{mirlin}. \\
In Figs. \ref{locali} and \ref{locali1} it is shown that $\lambda^\mathrm{ins}_{\mathrm{c}}$ decreases with $|U|$.
This enhancement of localization effects caused by attractive interaction is consistent with previous results in the literature \cite{shepe}. It is also observed that for $U = -1$ the dependence on $\lambda$ is not monotonous. Initially it decreases with $\lambda$ but close to the transition ($\lambda \lesssim \lambda^\mathrm{ins}_{\mathrm{c}}$) has a sharp peak before a steep drop right at $\lambda^\mathrm{ins}_{\mathrm{c}}$. 
This is not expected as it is believed that quasi long-range order is always weakened by disorder effects \cite{gosh}. Within a mean field approach this might be attributed to the enhancement of eigenstate fluctuations around the critical region \cite{condis}.
The absence of enhancement for larger $|U|$ is a consequence of the shorter coherence length in this case. Single particle fluctuations are suppressed if the coherence length becomes smaller than the system size.  \\
 We note that, according to Fig. \ref{locali1}, the transition to localization occurs at $\lambda^\mathrm{ins}_{\mathrm{c}} \approx 1.0$. On the other hand, according to Fig.\ref{stiffb} , global superfluidity breaks down at $\lambda_c \approx 0.95$. This suggests the existence of a metallic pseudo-gap phase for $0.95 < \lambda < 1.0$ characterized by strong pairing but no global superfluidity. We note the range of $\lambda$'s for which we observe this phase is relatively narrow and it seems to decrease for larger $U$. Therefore
we cannot discard the possibility that this metallic phase is a finite size effect, namely, the system is already an insulator but the localization length is larger than the system size. 
\subsection{Low energy excitations: Spin gap}
Finally we study the low energy excitations of Eq.(\ref{eqn:Hubbard}) by computing the minimum energy to break a pair, the so-called spin gap,
\be
\label{deltas}
\Delta_\mathrm{S} \equiv E_0(N+1, N-1) - E_0(N, N),
\ee
where $E_0(N_\uparrow, N_\downarrow)$ is the ground state energy for $N_\uparrow$ spin-up and $N_\downarrow$ spin-down atoms.
In Fig. \ref{spingap} we present results for $\Delta_{\mathrm{S}}$ for a fixed $L$ as a function of $\lambda$ and different $U$'s.
It is observed that $\Delta_{\mathrm{S}}$ is an increasing function of $\lambda$.
By contrast in 2D weakly disordered systems the gap decreases with $\lambda$ \cite{gosh} since the spectral density around the Fermi energy decreases with disorder. In quasiperiodic systems the situation is different.
 As $\lambda$ increases, the spectral density around the Fermi energy develops gaps at different scales and the spectral density in the remaining bands becomes higher. For not too large $\lambda$'s it is likely that, on average, there are no gaps around the Fermi energy. Therefore both the spectral density and the spin gap will increase as $\lambda$ increases.
Close to the metal-insulator transition, strong density-density fluctuations in the one-body problem \cite{condis,mirlin} further enhance the gap. This enhancement is a coherent effect and therefore it is expected to diminish as the coherence length becomes of the order of the system size which occurs in the region of strong coupling. 
For larger $\lambda$, already in the insulator region, the spin gap $\Delta_{\mathrm{S}}$ still increases with $\lambda$. This is not related to superconductivity but rather to the fact that now the gap is related to the mean level spacing which in the insulating region increases with disorder \cite{condis}.
\section{Conclusions}
We have studied the stability of the superfluid state in a 1D interacting and disordered Fermi gas. We have shown attractive interactions enhance localization effects. For intermediate couplings $|U| \approx 1$ we have identified a region close to the insulator transition in which superfluidity is substantially enhanced. Moreover our numerical results suggest that 
the breaking of global superfluidity might occur at a slightly weaker disorder than the insulator transition. If this is confirmed, a ``pseudo-gap'' metallic region characterized by pairing but no global superfluidity occurs between the two transitions. These results provide a theoretical framework for experimental studies of quantum phase transitions in 1D cold Fermi gases.\\

\acknowledgments
We thank Masahito Ueda for valuable conversations and a critical reading of the manuscript. 
Part of the computation has been done using the facilities of the Supercomputer Center, Institute for Solid State Physics, University of Tokyo. M.T. was supported by a Research Fellowship of the Japan Society for the Promotion of Science (JSPS) for Young Scientists. A.M.G. acknowledges financial support from DGI through Project No. FIS2007-62238 and from the JSPS. A.M.G. thanks Masahito Ueda and his group for their warm hospitality during his stay in the University of Tokyo.

\end{document}